# Some New Results and Perspectives Regarding the Kuiper Belt Object Arrokoth's Remarkable, Bright Neck

S. Alan Stern
Southwest Research Institute

Brian Keeney
Southwest Research Institute

Kelsi Singer
Southwest Research Institute

Oliver White
SETI Institute

Jason D. Hofgartner
Jet Propulsion Laboratory,
California Institute of Technology

Will Grundy
Lowell Observatory

and

The New Horizons Team

# Abstract

One of the most striking and curious features of the small Kuiper Belt Object (KBO), Arrokoth, explored by New Horizons is the bright, annular neck it exhibits at the junction between its two lobes. Here we summarize past reported findings regarding the properties of this feature and then report new results regarding its dimensions, reflectivity and color, shape profile, and its lack of identifiable craters. We conclude by enumerating possible origin scenarios for this unusual feature. New results include a new estimated measurement of the observed neck area of 8±1.5 km², a total neck surface area of 32 km², a 12.5:1 ratio of circumference to height, a normal reflectance histogram of the observed neck, and the fact that no significant (i.e., >2σ) neck color units were identified, meaning the neck's color is generally spatially uniform at the 1.5 km/pixel scale of the best color images. Although several origin hypotheses for the bright material in the neck are briefly discussed, none can be conclusively demonstrated to be the actual origin mechanism at this time; some future tests are identified.

## Introduction and Summary of Existing Knowledge

On 1 January 2019, the Kuiper Belt Object (KBO) (486958) Arrokoth (i.e., 2014 MU$_{69}$) became the first small KBO to be explored by any spacecraft, when NASA's New Horizons spacecraft flew past it, conducting numerous observations to reveal its properties (Stern et al. 2019). Arrokoth, discovered in 2014 by the Hubble Space Telescope (Buie et al. 2018), has an orbit that places it among the Cold Classical KBOs (CCKBOs). CCKBOs are a primordial population of bodies formed at or very close to their current heliocentric distance that they currently reside (e.g., Parker & Kavelaars 2010; Nesvorny 2018). These factors indicate that Arrokoth is the most dynamically and thermally primitive body ever explored by spacecraft (e.g., Moore et al. 2018; Stern et al. 2018).

New Horizons made its closest approach (CA) to Arrokoth on 1 January 2019 at 05:33:22.4 (±0.2 seconds, 1σ) universal time (UT). The CA distance, 3538.5±0.2 (1σ) km, was to the celestial north of Arrokoth; the spacecraft's speed past Arrokoth was 14.43 km s$^{-1}$. The approach angle was 11.6 deg from the vector to the Sun.

New Horizons carries a suite of seven scientific instruments (see the collected instrument review papers in Russell et al. 2008). These are: (i) Ralph, consisting of two components, the Multispectral Visible Imaging Camera (MVIC), a multicolor/panchromatic mapper, and the Linear Etalon Imaging Spectral Array (LEISA), an infrared composition mapping spectrometer; (ii) the Long Range Reconnaissance Imager (LORRI), a long focal length panchromatic visible camera; (iii) the Alice extreme/far ultraviolet mapping spectrograph; (iv) a Radio Experiment (REX) to measure surface brightness temperatures and X-band radar reflectivity; (v) the Solar Wind Around Pluto (SWAP) charged particle solar wind spectrometer; (vi) the PEPSSI (Pluto Energetic Particle Spectrometer Science Investigation) MeV charged particle spectrometer; and (vii) the Venetia Burney Student Dust Counter (VBSDC)—a dust impact detector.

New Horizons data from the flyby of Arrokoth revealed it to be a 36 km long, moderately dark, red, contact binary. The measured visible color slope and a 1.25-2.50 μm reflectance spectrum revealed a surface at least partially comprised of methanol ice ($CH_3OH$) mixed with some unidentified refractory organic material (Stern et al. 2019; Spencer et al. 2020; Grundy et al. 2020). The measured rotation period of Arrokoth is 15.9 hours (Stern et al. 2019; Spencer et al. 2020). Clues from Arrokoth's geology, shape, and contact binary nature point to it being created by the low velocity merger of two bodies formed in a local collapse cloud streaming (or similar) instability (Stern et al. 2019; McKinnon et al. 2020; Marohnic et al. 2021).

One of the most prominent and curious features on Arrokoth's surface is its bright "neck," located at the interface of its two merged lobes (seen in Arrokoth Close Approach observation 6 (CA06) in Figure 1, the highest resolution observation of Arrokoth). The neck occupies an annular, approximately azimuthally symmetric region at the contact between Arrokoth's two lobes. Bi-lobate contact binaries appear to occur frequently in the Kuiper Belt (e.g., Thirouin & Shepard 2019), however the bright material in Arrokoth's neck has no direct parallel feature yet detected in the Solar System; perhaps new imaging of KBOs visited by spacecraft will find more such features.

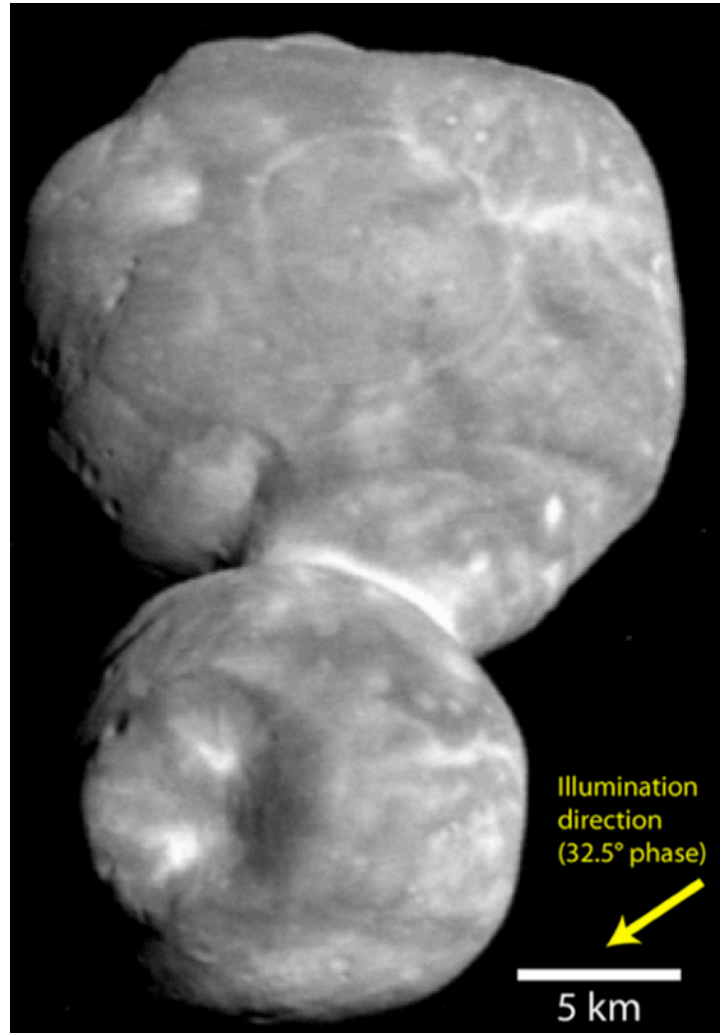

**Figure 1. Deconvolved and stacked Arrokoth close approach LORRI mosaic CA06 obtained by New Horizons at range 6,650 km, solar phase 32.5°, and a pixel scale of 33 m/pixel. Arrokoth's neck is the bright, annular region at the juncture between its larger and smaller lobes.**

Hofgartner et al. (2021) showed that Arrokoth's neck is significantly more reflective[1] than the average across its surface, i.e., the average normal albedo of the neck is 0.365±0.033 (1σ) compared to an Arrokoth average of 0.24, respectively. That result was obtained using the New Horizons Long Range Reconnaissance Imager (LORRI) camera's, with a 0.61 µm pivot wavelength (Weaver et al. 2008); Hofgartner et al. (2021) describe their technique for measuring this reflectance in detail.

---

[1] **Here "reflectance" means normal reflectance, the I/F corrected with a model to normal incidence and emission (see Hofgartner et al. 2021). Normal reflectance is sometimes also called normal albedo.**

The neck has a diffuse margin at least on the large lobe side, but extreme foreshortening of New Horizons images due to the spacecraft-KBO relative geometry at flyby makes it difficult to characterize whether this diffuse margin is also exhibited on the small lobe side (Spencer et al. 2020).

Grundy et al. (2020) used the Ralph instrument Blue (400-550 nm), Red (540-700 nm), and NIR (860-910 nm) filter data to compute the average observed visible color reflectance slope of Arrokoth at 27.5±1.5% (1σ) per 1000 Å at 5500 Å; the data used for this computation were at a solar phase angle of 41 deg. As noted in that paper, subtle color differences exist, as several other features on Arrokoth's surface are less red than its average color, with visible color slopes as low as 23±2% per 1000 Å. The neck itself has a visible wavelength color slope of 25±1% per 1000 Å. Globally, we note that KBOs exhibit a broad range of optical bandpass special slopes from values of a few to over 30% per 1000 Å; binary CCKBOs however exhibit optical spectral slopes of between ~15% and 45% per 1000 Å, placing Arrokoth near the middle of this broad range of spectral slopes (Fraser et al. 2020).

New Horizons revealed that the absorption signature of methanol ($CH_3OH$) is found in the neck region (Stern et al. 2019; Grundy et al. 2020), as it is as well is across the object, but SNR limitations prevent significantly constraining any difference in methanol abundance at the neck vs. elsewhere on Arrokoth. Aside from the methanol and complex organics (e.g., tholins, e.g., Cruikshank et al. 2005) inferred from Arrokoth's albedo, color, and spectral model, no other surface constituents were found in the neck region or elsewhere on Arrokoth (Grundy et al. 2020).

McKinnon et al. (2020) calculated that an Arrokoth bulk density >0.25 g cm$^{-3}$ would imply that the neck region is in compression, and that an Arrokoth density >1.25 g cm$^{-3}$ would cause the neck to collapse in compression if Arrokoth's cohesive strength is low and comet-like (i.e.,

<10 kPa; Groussin et al. 2019; Holsapple & Housen 2007). Spencer et al. (2020) noted that neither distortion nor faults are seen on the neck at available imaging resolution, suggesting that the density is not near the 1.25 g cm$^{-3}$ limit and also that merger speeds of the two lobes must have been very low.

It has also been noted (Spencer et al. 2020) that Arrokoth's bulk density must be ≥0.29 g cm$^{-3}$ because mutual gravity would otherwise be less than the centrifugal acceleration at Arrokoth's measured spin rate. Spencer et al. (2020) also found that local acceleration slopes (i.e., the gradients of the gravitational and rotational potential across the surface) are low on Arrokoth except in the neck, where they can exceed 35°.

Regarding gravitational geopotential, McKinnon et al. (2020) found that it reaches a global minimum at the neck. For a reference bulk density of 0.5 g cm$^{-3}$, surface gravitational slopes are downward into the neck region. Thus if loose material were to be generated (for example by impacts or thermal stresses) and then to subsequently flow downslope, some of it will naturally tend to collect in the neck region.

Regarding thermal conditions at the neck, Grundy et al. (2020) showed that, owing to self-shadowing, the neck region generally receives less energy than Arrokoth's equatorial zone, but that the radiation received at the neck from thermal emission by other parts of Arrokoth near the neck causes the neck to reach higher annual maximum temperatures, near 60 K, than almost any other part of the KBO.

### Newly Derived Neck Properties

Now we present a series of new results regarding the dimensional, morphological, reflectance, and cratering properties of Arrokoth's neck; we take each in turn.

*Neck Dimensions and Morphology.* The diameter of the neck is ~7±1 km as measured using the CA06 observation. We fit an elliptical ribbon (see Figure 7) around the observed portion of Arrokoth's neck using the shape model of Spencer et al. (2020) to best enclose the bright pixels in the neck. We found the areal extent of the observed neck is 8±1.5 km², indicating that if the neck continues onto Arrokoth's far side, as we expect, its total surface area is of order 20 km². However, the partial eclipsing of the neck by the foreground small lobe almost certainly makes our measurement an underestimate.

Using observation CA06, the short dimension of the neck (i.e. its 'height') measures 0.9±0.1 km. However, an observation that offers a very different view of the neck from that of CA06 and other approach imaging is the high phase CA07 LORRI observation, obtained 9.4 minutes after closest approach, which shows the illuminated double crescent of Arrokoth (see Figure 2a). In contrast to the approach images, the large lobe is in the foreground and the small lobe is in the background here, with the neck now seen at an angle nearly orthogonal to that observed in the approach observations. The neck region measures ~9 pixels across, and we consider it to be resolved to a degree that is sufficient to confirm the "U" shape as genuine and not merely a poorly resolved V-shape. Crucially here, its profile does not appear to be eclipsed by either lobe. Figure 2b annotates that limb profile of Arrokoth with sections of the limb interpreted to be those of the large and small lobes and the neck in-between, the profile of which has a noticeable "U" shape that is not apparent in the frontal views of the approach images, in which the eclipse of the neck by the small lobe causes the limb to pinch at a sharp, convex "V" (see Figure 1).

We point out that in these very high solar phase angle observations, the brightest pixels are likely the result of the extreme illumination conditions rather than the intrinsic albedo of the surface. Hence, we do not interpret the bright spot just to the left of the neck in CA07 (see yellow arrow) to be part of the neck, but rather a brief continuation of the profile of the large lobe in the foreground. We also see that the neck region (in red) terminates where the curvature of its profile starts to flatten out on either side of the "U." Using this definition, the width of the neck region here is ~1.6±0.2 km, a larger value than that measured from CA06, and which provides an indication of the fraction of the neck that is eclipsed by the small lobe in CA06. Adopting that value for the neck height would suggest a total neck surface area of 32±5.5 km² and a fiducial 12.5:1 ratio of circumference to height.

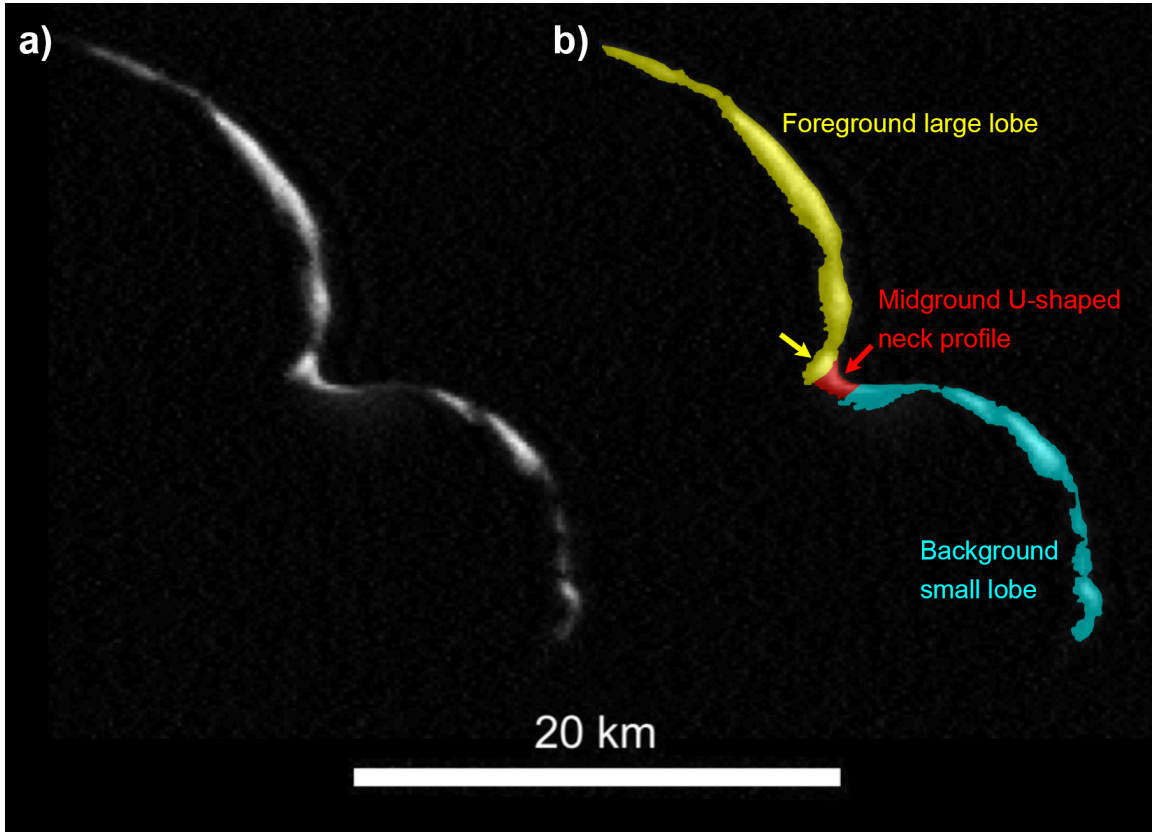

**Figure 2.** Original (panel a) and annotated (panel b) versions of the LORRI CA07 observation of Arrokoth, obtained at a solar phase angle of 152.4° and a pixel scale of 175 m. In both cases, the image has been deconvolved to remove motion smear. The portions of the limb that are interpreted to correspond to the large lobe, small lobe, and neck are highlighted by different colors in panel b. The yellow arrow points to a location where the limb profile appears to subdivide, with a short protrusion extending to the lower left representing the termination of the limb of the foreground large lobe, and the limb of the neck extending to the right.

We also point out that the "U" shape of the neck profile seen in CA07 transitions smoothly into the profile of the small lobe, with no apparent break in slope between the two in the 175 m/pixel imaging, which is consistent with the neck being filled with loose material that has sloughed off both of the lobes to collect within the gravitational low of the neck.

*Normal Reflectance of the Neck*. Figure 3 shows the normal reflectance distribution of Arrokoth's neck at 0.61 µm compared to both the normal reflectance distribution for the whole of Arrokoth (i.e., all pixels on the encounter hemisphere, including the neck) and that of other bright curvilinear features on Arrokoth. Figure 4 depicts the regions included as part of the neck and bright curvilinear features. The normal reflectance was determined from the best-fit lunar-Lambert photometric function of Arrokoth in Hofgartner et al. (2021). The Hofgartner et al. (2021) results are in turn based on the shape model of Spencer et al. (2020); illumination conditions are described in Hofgartner et al. (2021). From Figure 3 we find that the neck's normal reflectance is distinctly higher than Arrokoth as a whole. Further, Arrokoth's bright curvilinear features are not as bright as the neck. However, there is some overlap between all three distributions as can be seen in Figure 3. All three normal reflectance distributions are seen to be unimodal. The individual distributions of the bright curvilinear features are shown in Figure 5. This comparison reveals that although the distribution of normal reflectance values has a similar width, the neck is brighter than other curvilinear features on the KBO. Note, however, that there are bright spots on Arrokoth with maximum normal reflectance values that are comparable to the maximum normal reflectance of the neck (see also Hofgartner et al., 2021).

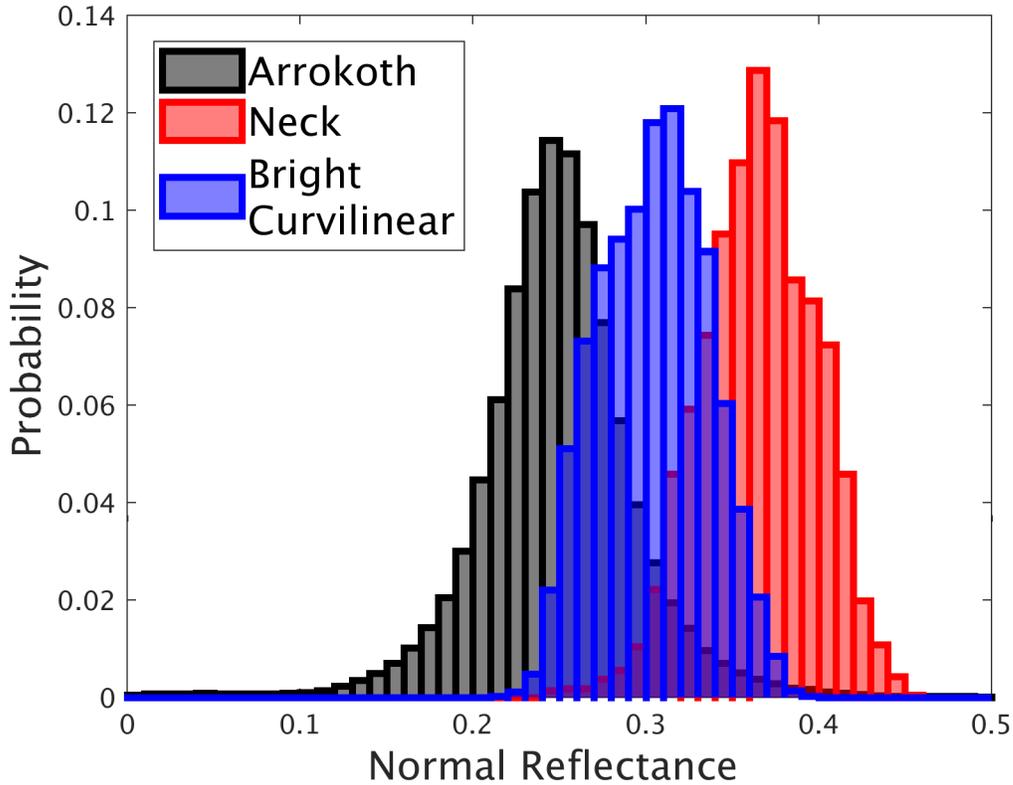

**Figure 3.** Normal reflectance distributions of all pixels on Arrokoth, its neck, and bright curvilinear features (see Figure 4) at the New Horizons LORRI pivot wavelength of 0.61 μm. The probability distribution here is the number of pixels in a bin divided by the total number of pixels for that feature. The total number of pixels are given in Table 1 below.

Table 1 provides the mean normal reflectance and standard deviations for the regions used in Figures 3 and 5.

**Table 1: Normal Reflectance Values and Standard Deviations for Various Neck Regions**

|  | Number of Pixels | Normal Reflectance Mean | Normal Reflectance Standard Deviation |
|---|---|---|---|
| **Neck** | 2112 | 0.365 | 0.033 |
| **BCLs (All)** | 7368 | 0.305 | 0.031 |
| **BCL1** | 1238 | 0.316 | 0.018 |
| **BCL2** | 2559 | 0.298 | 0.028 |
| **BCL3** | 1828 | 0.279 | 0.022 |
| **BCL4** | 1743 | 0.333 | 0.021 |

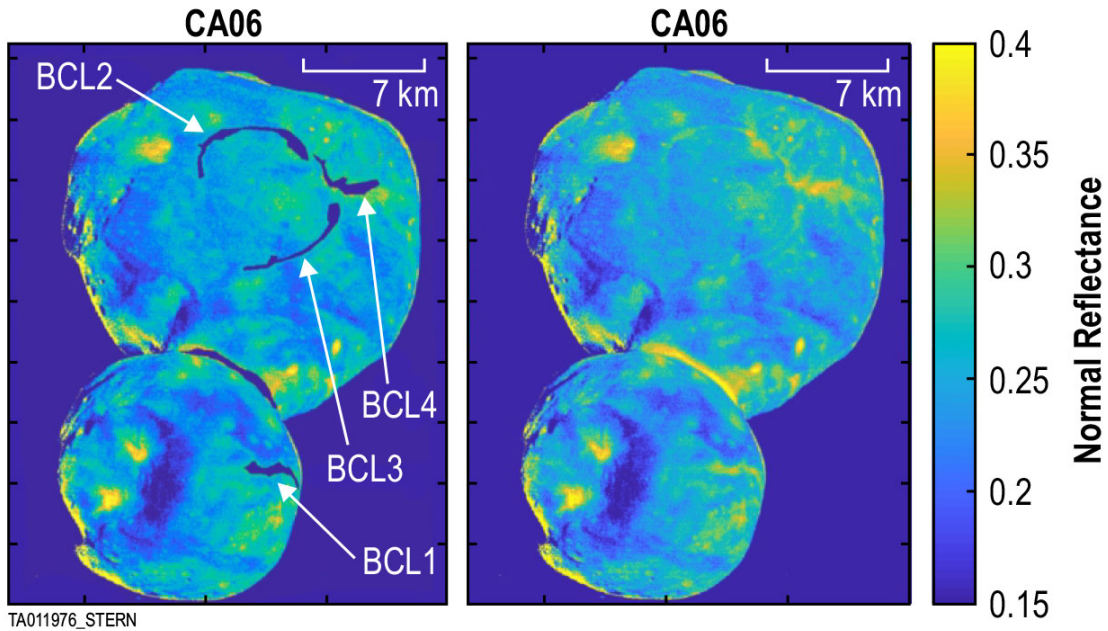

Figure 4. A normal reflectance map of Arrokoth at the New Horizons LORRI pivot wavelength of 0.61 μm, from Hofgartner et al. (2021). Pixels included in the normal reflectance histograms of the neck and Bright Curvilinear (BCL) features in Figure 3 are indicated by the areas masked in dark blue. Panel (a): Annotated; Panel (b) Unannotated.

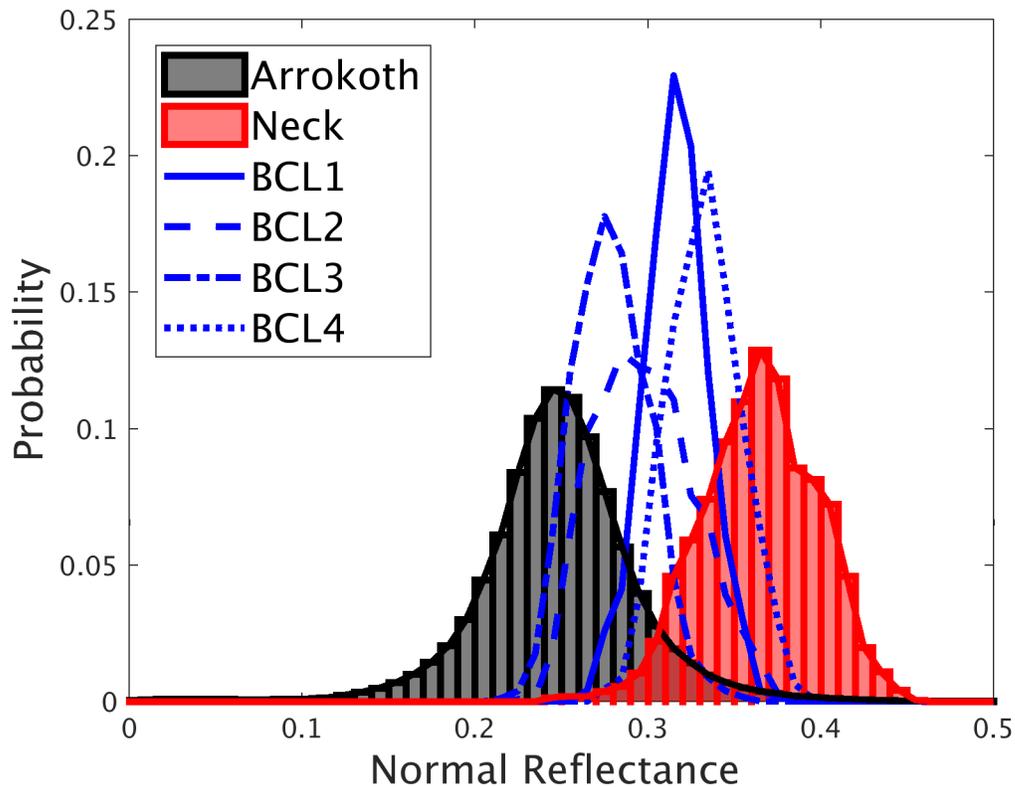

**Figure 5. Normal reflectance distributions of the four individual Bright Curvilinear (BCL) features we analyzed, again compared to the neck and all pixels on Arrokoth as a whole. Their combined distribution is in Figure 3. The histograms are shown as line plots so that it is easier to distinguish between them.**

*Reflectance Variations on the Neck.* Figure 6 shows I/F spatial contours on the neck, drawn at 1, 2, 3, and 4σ above the I/F mean of this image. For the pixels shown in Figure 6, the mean I/F is 0.056±0.010, uncorrected for any disk function; the mean statistical uncertainty in I/F is ~4× smaller than the contour spacing in Figure 6. From this we conclude that the contours reveal statistically meaningful differences in measured reflectance; but whether this is due to changes in albedo or slope effects is unclear. From this figure we also find that the lowest contour reveals a skirt of material bordering the large lobe side of the neck. A similar skirt may well exist on the small lobe side, but projection effects due to the spacecraft-KBO geometry around closest approach prevent us from confirming if this is so.
13

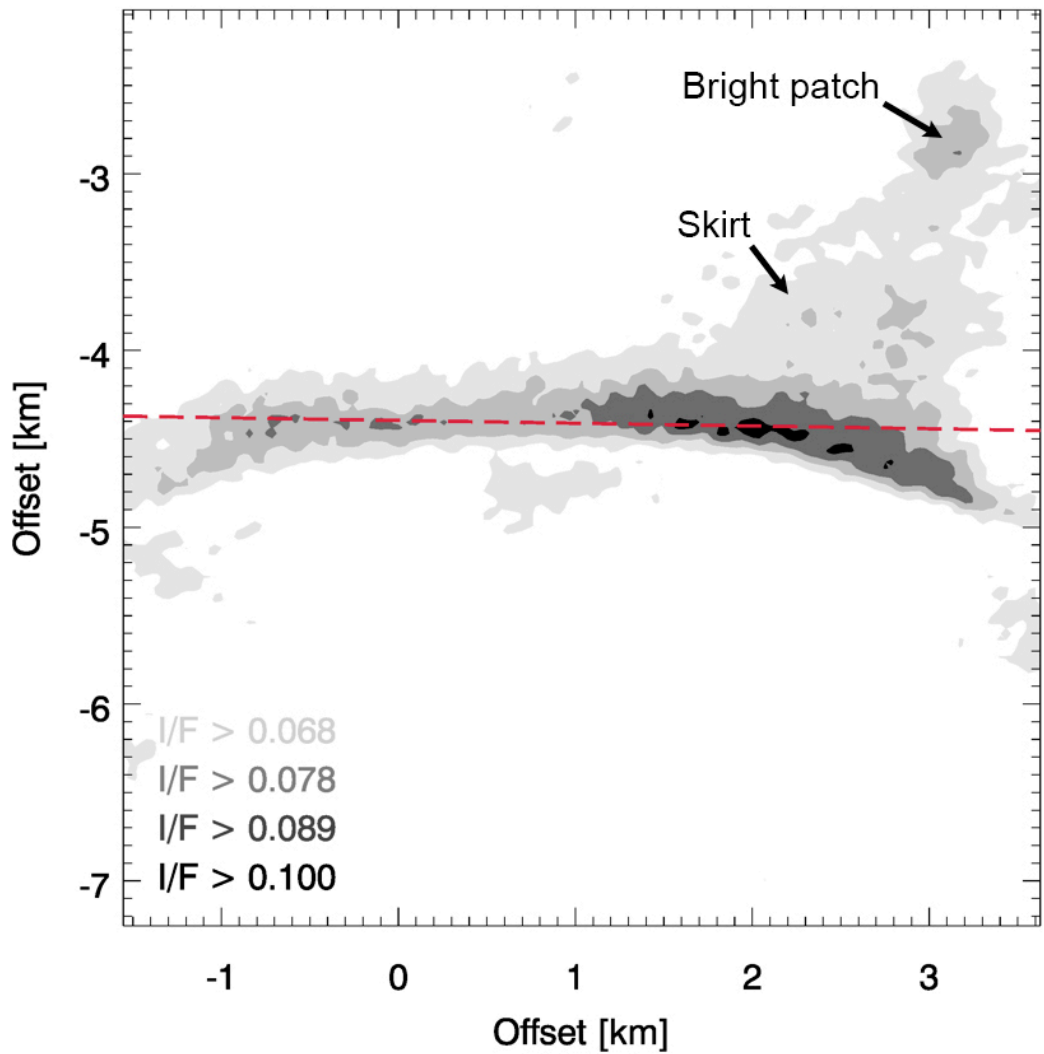

**Figure 6. I/F contours on the neck at 1, 2, 3, and 4σ above the I/F mean. The axes measure the offset from the CA06 Arrokoth image center. The red dashed horizontal line illustrates that there is observed curvature even when the neck is rotated to be "horizontal."**

To investigate neck reflectance structures further, we subdivided the neck into 17 approximately equal areas around its CA06 observable circumference, each with a surface area of 0.5 km². Figure 7 contains the average I/F cut across each of these 17 areas along their normal direction to the neck's azimuthal axis. The error bars shown reflect 1σ counting statistics. The red curves under each of these neck I/F cuts show the full width at half maximum (FWHM) of a sharp edge (specifically, Arrokoth's distal limb) revealing the 2.4 pixel spatial scale at which albedo markings can be reliably resolved. Figure 7 demonstrates that both in terms of the counting statistics and the image spatial resolution, real I/F structures are seen in almost every cut across the neck's short dimension. These may relate to actual intrinsic albedo variations, slope variations at scales we cannot resolve, varying effects of surface roughness and porosity, or some combination of all of these factors. The lack of discernible km-scale stereo topography on the neck (Schenk et al. 2021; Schenk pers. comm.) is unfortunate, preventing us to better resolve this ambiguity.

We do note, however, a possible geological relation whereby the skirt of material on the large lobe side of the neck appears to terminate at a sub km-sized bright patch (see Fig. 6) which may represent a pit (likely an impact crater; Spencer et al. 2020) filled with loose, bright material. Given that the pit exists at a location where gravitational slopes are amongst the highest encountered on Arrokoth (>20°; McKinnon et al. 2020), loose material collecting there may be particularly disposed to overflow and then migrate downslope towards the neck, e.g., if mobilized by the shock of impacts occurring elsewhere on the surface. As such this skirt of material, as well as the particularly bright segment of the neck that it connects to. may represent a surplus of loose, bright material that has migrated from this pit. Alternatively, this small, bright patch of material on the skirt may just be in its own local topographic low/gravitational pocket, collecting there originally as typical other bright material on Arrokoth has.

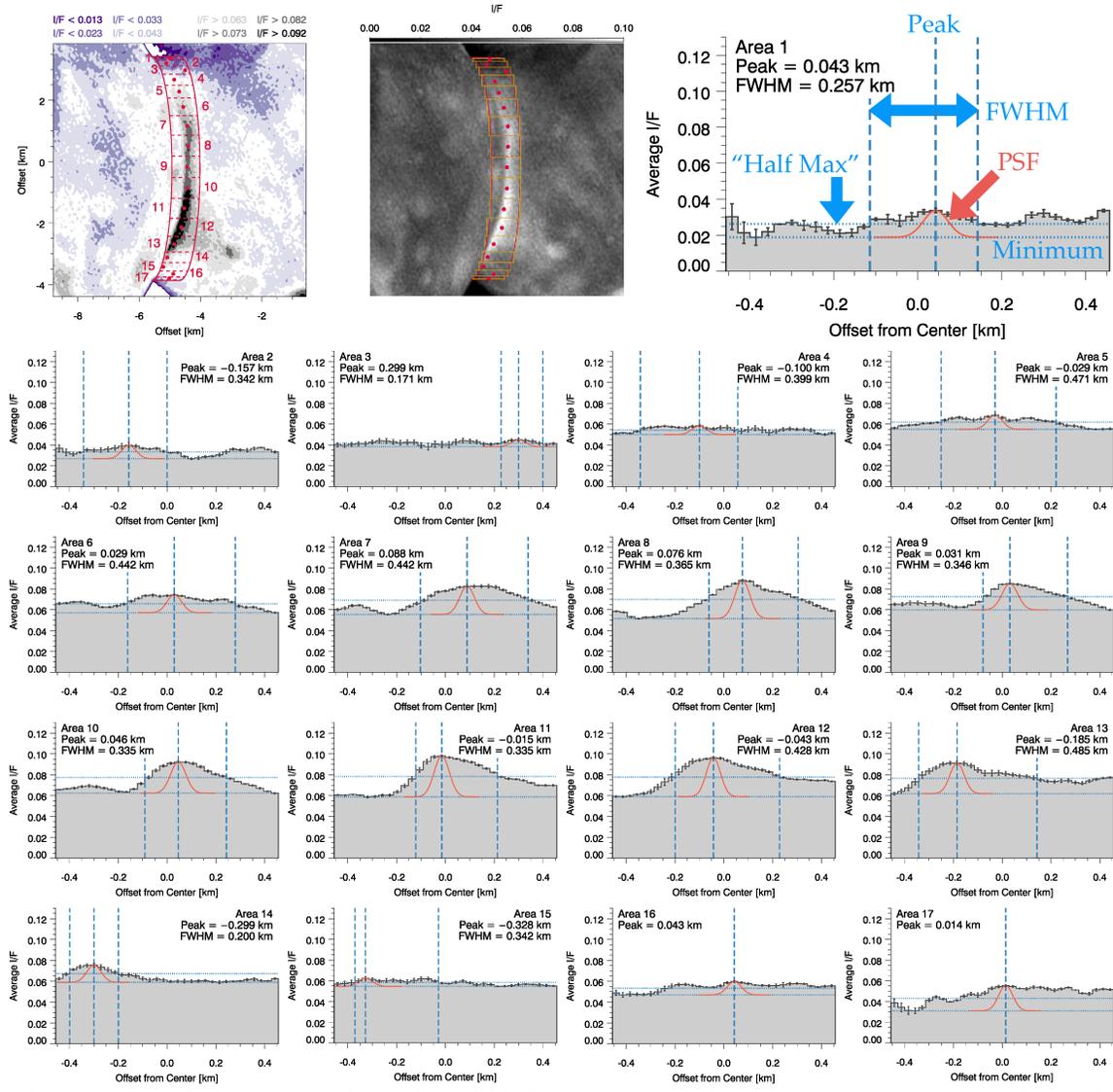

**Figure 7.** Plots showing 17, 0.9 km-wide I/F cuts across Arrokoth's neck along with context imaging in the upper left. Note that the FWHM values shown by the blue vertical lines in each panel are wider than the red point spread function (PSF) widths.

*Color Variations on the Neck.* Using methods very similar to Grundy et al. (2020), we used MVIC color ratios of all pixels seen on the neck to re-derive and confirm the Grundy et al. (2020) result that Arrokoth's neck is slightly less red than the KBO's average. The spectral slope of the neck (specifically, pixels inside the elliptical ribbon) at 5500 Å is 25.4±2.0% (1σ) per 1000 Å, and that of the surroundings is 27.7±2.2% (1σ) per 1000 Å, which are within the error bars indistinguishable from the Grundy et al. (2020) results. We also created color ratio maps of the neck, but no significant (i.e., >2σ) color variations were identified, meaning the neck's color is generally spatially uniform in color at the 1.5 km/pixel scale of the best color images.

*A Cratering Search on the Neck.* We carefully inspected all the resolved LORRI panchromatic images of Arrokoth to search for evidence of craters on the neck. Such craters could serve to constrain the neck's age and the depth of its bright albedo surface. However, no craters were found, even in the highest resolution images at 33 m/pixel. Is this surprising? Figure 8 presents cratering results from Spencer et al. (2020), displayed as the differential number of craters per size and unit area. The data is displayed for different subsets of the features on Arrokoth. Maryland is the informal name given to the one large (~7 km diameter) crater on the small lobe. The designations "high," "medium," and "low" indicate how confident a team of crater mappers was that any given feature was in fact an impact crater (based on the morphology of the feature). The prefix "A" denotes features across all of Arrokoth, whereas "LL" indicates features only on the large lobe. The large lobe was further broken down into the terminator region (Term) and the features that appeared as bright spots (Bright) vs. those with discernable negative topography (Pits). These different subsets were investigated in Spencer et al (2020) to examine the range of possible crater areal densities across the body. Combining the visible dimensions of the neck given above with Figure 8 data that exclude terminator terrains, we estimate the number of resolvable craters one should expect to find on the observed area of the neck to be of order ~0.1-1. Therefore, the lack of observable craters on the neck is not a significant constraint on age.

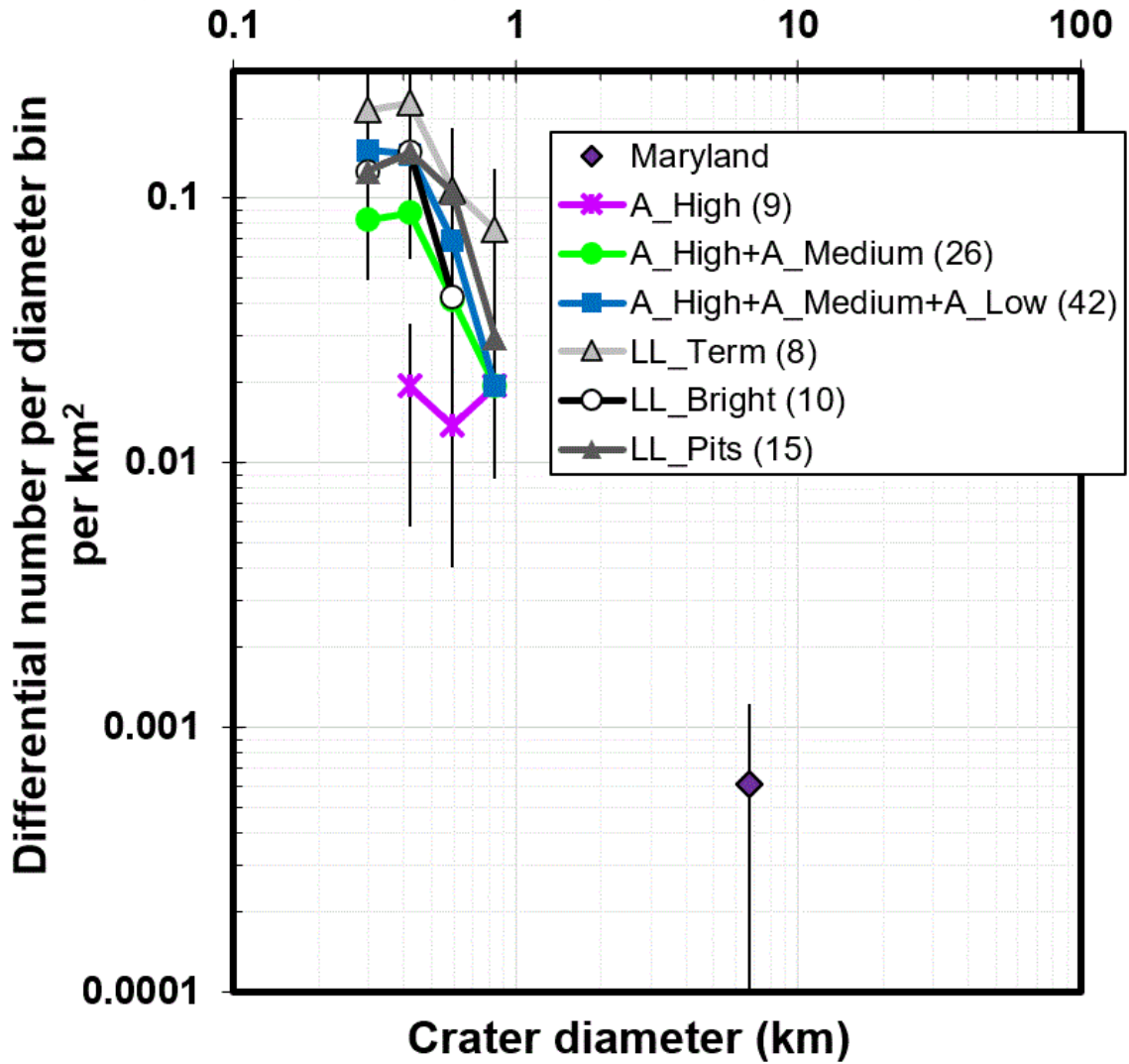

**Figure 8. Differential number of craters per unit area vs. crater diameter for various terrains on Arrokoth. Adapted from Spencer et al. 2020; see text for details of the subgroups.**

**Regarding the Bright Neck's Origin**

As pointed out above and in both Stern et al. (2019) and Spencer et al. (2020), the numerous bright curvilinear features on Arrokoth share somewhat similar albedos and colors to the neck; they also share the attribute of being found in topographically low lying areas. But recall that these other features are not quite as bright as the neck (Figures 4 and 5). Therefore, it is reasonable to suggest that the neck could be the largest scale example of a general process that creates or aggregates bright and preferentially bluer low-lying material across Arrokoth. A similar effect was noted on the surface of comet 67P/Churyumov-Gerasimenko which was examined in detail by the Rosetta mission (e.g., Fornasier et al. 2019). In this scenario, the more striking appearance of the neck than the lineations could, for example, be related to its more extreme topographic and higher gravitational slopes. The "U" shape of the neck when seen in profile at high solar phase angle is consistent with (but does not definitively require that) loose material accumulated there, as otherwise a sharper angle at the contact between the limbs of the two lobes could be expected.

A mechanism that may have mobilized small particles involves the early loss of supervolatiles such as CO or $CH_4$, which were probably frozen as ices onto the outer nebular grains that Arrokoth accreted from (Grundy et al. 2020). After the nebular dust cleared and the Sun began to illuminate Arrokoth's surface, its interior would have gradually warmed, driving sublimation loss of such materials (e.g., Zhao et al. 2020; Lisse et al. 2021; Steckloff et al. 2021). The resulting gaseous outflow could have mobilized materials enabling down-slope movement. This escaping gas could also have sorted particles by size and density, providing one explanation for the distinct scattering properties of material now in Arrokoth's topographic lows, though it is puzzling that such a bright albedo feature could persist (e.g., against space weathering and meteoritic bombardment) over 4+ Gyr, though a dramatically lower age for the bright neck poses other difficult to resolve problems.

Intriguingly, two bright spots in the large crater on the small lobe, as well as two bright spots in low areas on the large lobe near the neck, are of comparable brightness to the neck. And yet, the darkest regions of Arrokoth also occur in depressions (Hofgartner et al. 2021). Albedo and topography may be related and may influence one another on Arrokoth, but the details of such a relationship are unclear. Improved shape and topography as well as photometric analyses and their correlations may reveal further clues.

However, the origin of the neck feature at the merger zone of Arrokoth's two lobes might not have the same origin as the non-neck bright locales on the KBO. Some bright neck origin possibilities could include: (i) the merger impact extrusion of preexisting bright interior icy material on or inside one or both lobes, (ii) the thermal extrusion of bright material following merger due to changes in thermal properties or conditions (such as reduced radiative cooling) at the merger zone subsequent to lobe-on-lobe contact (Stern et al. 2019), (iii) evolutionary processes such as reduced space weathering (due to the reduced solid angle visible at the neck to sky radiation) or enhanced thermal effects (e.g., due to re-radiation from the two lobes on either side of the neck; Binzel et al. 2019; Spencer et al. 2020), (iv) merger-related tidal forces concentrating any bright, mobile, fine material there analogous to a scree, and (v) sublimation driven volatile transport across Arrokoth to the neck (Katz & Wang 2021).

Unfortunately, tests that might discriminate between these various scenarios (e.g., using the composition and microphysical properties of the neck compared to adjacent terrains) seem impossible without higher resolution imaging, photometry, and spectroscopy of Arrokoth, or perhaps another KBO displaying a similar feature. Since such prospects are at least decades off, the only tools available in the nearer term are likely to be numerical modeling, laboratory studies (particularly of regolith size sorting mechanisms), and possibly active experiments on bilobate small bodies with similar surface gravity much closer to Earth.

## Acknowledgements

We thank the NASA New Horizons project for support. Part of this research was carried out at the Jet Propulsion Laboratory, California Institute of Technology, under a contract with NASA. We also thank two anonymous referees for their work to improve and clarify this paper.20